\documentclass[10pt]{JHEP3}
\usepackage{amssymb}
\usepackage{amsmath}
\usepackage{bm}

\title{Primordial SdS universe from a 5D vacuum: scalar field fluctuations on Schwarzschild and Hubble horizons}
\author{Jos\'{e} Edgar Madriz Aguilar,\\
Instituto de F\'isica de la Universidad de Guanajuato, C.P. 37150,
Le\'on Guanajuato, M\'exico. E-mail:
\email{jemadriz@fisica.ugto.mx}}
\author{Mauricio Bellini\\
Departamento de F\'isica, Facultad de Ciencias Exactas y
Naturales, Universidad Nacional de Mar del Plata, Funes 3350, C.P.
7600, Mar del Plata, Argentina.\\
Instituto de Investigaciones en F\'{\i}sica, Consejo Nacional de
Investigaciones Cient\'ificas y T\'ecnicas (CONICET), Argentina.\\
E-mail: \email{mbellini@mdp.edu.ar, mbellini@conicet.gov.ar}}

\abstract{We study scalar field fluctuations of the inflaton field
in an early inflationary universe on an effective 4D
Schwarzschild-de Sitter (SdS) metric, which is obtained after make
a planar coordinate transformation on a 5D Ricci-flat
Schwarzschild-de Sitter (SdS) static metric. We obtain the
important result that the spectrum of fluctuations at zeroth order
is independent of the scalar field mass $M$ on Schwarzschild
scales, while on cosmological scales it exhibits a mass
dependence. However, in the first-order expansion, the spectrum
depends of the inflaton mass and the amplitude is linear with the
Black-Hole (BH) mass $m$.} 
\begin{document}
\maketitle

\section{Introduction}

The cosmic microwave background that we observe today is almost
isotropic. The background temperature is about 2.7 K with a tiny
fluctuation at a level of about $10^{-5}$ k. This is consistent
with measurements of matter structures in the universe at
cosmological scales, where the universe is almost homogeneous.
However, it is well known that the mass spectrum on astrophysical
and galactic scales has a negative index which decreases with the
observed scale. The studies developed in the last years have shown
that on astrophysical scales the power spectrum of galaxies and
clusters of galaxies can be satisfactorily expressed by a power
law with an index between $-1.9$ and $-1.5$ \cite{einasto}. On
larger scales the spectrum turns over reaching a maximum on scales
of $(100 - 150) \ h^{-1} \ Mpc$. For our purposes we shall assume
that the evolution of the
structure in the universe is only due to gravity.\\

Astrophysical and Cosmological systems have been also
theoretically investigated, in the context of theories of gravity
in more than four dimensions (4D). In theories with five
dimensions (5D), models fit the observations due in much because
they smoothly embed 4D solutions in 5D solutions \cite{Wbook}.
Since the M-theory and quantum geometry have appeared as promising
candidates for a quantum theory of gravity \cite{Maartens}, the
idea that our universe is a 4D space-time embedded in a higher
dimensional manifold with large extra dimensions, has been a topic
of increased interest in several branches of physics, and in
particular, in cosmology. This idea that goes back to the works of
Nordstr$\ddot{o}$m \cite{Nor} and the original version of the
Kaluza-Klein theory \cite{KKT}, has generated a new kind of
cosmological models, including the
type of quintessential and dark energy models.\\

In particular, theories regarding just one extra dimension have
become quite popular in the scientific community. Two proposals
that have been subject of great research are the brane theory
\cite{BT} and the induced matter (IM) theory of gravity
\cite{stm}. Although brane theory and IM theory have different
physical motivations for the introduction of a large extra
dimension, they are equivalent each other \cite{Equi}. They both
predict identical non-local and local high energy corrections to
general relativity in 4D, and usual matter in 4D
is a consequence of the metric dependence on the fifth extra coordinate \cite{Equi}. \\

In this letter we have particular interest in the ideas of the IM
theory. A success of this theory is that classical properties of
matter fields in 4D can be given a geometrical interpretation if
we assume a 5D apparent vacuum, defined by the condition:
$^{(5)}R_{ab}=0$. Matter fields in 4D are described by an
effective 4D energy-momentum tensor purely geometrical in origin
\cite{2}. In particular, according to \cite{Wbook} the mass of
classical particles in 4D is directly related with the existence
of the fifth extra coordinate. In this manner in the IM theory the
mass of classical particles has a geometrical origin and this
feature makes this theory attractive from the fundamentalist point
of view. Furthermore, the fact that any energy-momentum tensor can
be geometrically induced in this manner, is hardly supported by
the Campbell-Magaard theorem and their extensions \cite{CMT1}. A
particular and very interesting 1-body solution of the 5D field
equations of the IM theory is the 5D Ricci-flat Schwarzschild-de
Sitter (SdS) black-hole, which contains an induced cosmological
constant \cite{11}. This solution can be embedded into a brane
world by using a conformal factor \cite{12} and constitutes a 5D
Ricci-flat SdS black string space. This black string space is the
Randall and Sundrum \cite{rs} 2-brane model. We have recently
shown in \cite{antigravity} that this 5D SdS BH solution give us
in 4D ordinary gravitational solutions on small (planetary and
astrophysical) scales, but repulsive (anti-gravitational) forces
on very large (cosmological) scales. We called the effective 4D
gravitational theory derived from this scenario, an extended
version of general relativity. This behavior of gravity on
different length scales derived from  this 5D framework, leave us
to put on the desk the question weather repulsive gravity can be
considered as a candidate for explaining the present accelerated
expansion of the universe. However, as it is well known in the
literature, inflation is another period of accelerated expansion
that the universe has passed through. So it makes interesting to
ask about if inflation still works in this extended version of
general relativity. In order to address the former question, in
this letter we define a free scalar field on a 5D geometrical
background, on which the 5D metric is expressed in planar
coordinates in order to consider in 4D, a black hole in an
inflationary scenario. The original massless 5D scalar field can
be seen in 4D as a massive scalar field, that we identify with the
inflaton field. The fifth extra coordinate is responsible for the
5D massless scalar field to acquire mass in 4D. This way using
some ideas of the IM theory we can give a geometrical origin for
the inflaton mass in 4D, and this is one contribution of the extra
coordinate in 4D. Our goal in this letter is to study the
evolution of quantum fluctuations during inflation on cosmological
scales, on which gravitational effects are repulsive, and also on
very short scales (which are close to the Schwarzschild scale),
where gravity manifests itself as attractive. \\

The letter is organized as follows. In section I we give a brief
introduction. In section II we obtain the dynamical equations for
the inflaton scalar field in 5D and 4D. We then proceed to section
III, divided in several subsections, in which we calculate the
inflaton quantum fluctuations on short (Schwarzschild) scales and
large (cosmological) scales. We have implemented an expansion for
the inflaton field in powers of a small parameter that arises
naturally from the model, so we calculate the power spectrum on
short and large scales at zeroth and first orders in the expansion
only. Finally in section IV, we give some final comments.

\section{The 5D scalar field equation of motion}

We consider the 5D Ricci-flat metric \cite{antigravity}
\begin{equation}\label{a1}
dS_{5}^{2}=\left(\frac{\psi}{\psi_0}\right)^{2}\left[c^{2}f(R)dT^{2}-\frac{dR^2}{f(R)}
-R^{2}(d\theta^{2}+sin^{2}\theta d\phi^{2})\right]-d\psi^{2},
\end{equation}
where $f(R)=1-(2G\zeta\psi_{0}/Rc^2)-(R/\psi_0)^{2}$ is a
dimensionless function, $\lbrace T,R,\theta,\phi\rbrace$ are the
usual local spacetime spherical coordinates employed in general
relativity and $\psi$ is the space-like extra dimension. Following
the approach of the IM theory, $\psi$ is here on
considered as non-compact. This metric is a 5D extension of the 4D
SdS metric. In this line element $\psi$ and $R$ have length units,
$\theta$ and $\phi$ are angular coordinates, $T$ is a time-like
coordinate, $c$ denotes the speed of light, $\psi_0$ is an
arbitrary constant with length units and the constant parameter
$\zeta$ has units of $(mass)(length)^{-1}$. As we can see, this
metric is written on a static chart coordinate, so in order to get
this metric written on a dynamical chart coordinate $\lbrace
t,r,\theta,\phi \rbrace$, let us to use the coordinate
transformation given by \cite{planar}
\begin{equation}\label{a2}
R=ar\left[1+\frac{G\zeta\psi_0}{2ar}\right]^{2},\quad
T=t+H\int^{r}dR\,\frac{R}{f(R)}\left(1-\frac{2G\zeta\psi_0}{R}\right)^{-1/2},
\psi=\psi,
\end{equation}
being $a(t)=e^{Ht}$ the scale factor, and $H$ the Hubble constant.
Thus the line element (\ref{a1}) can be written in terms of the conformal time
$\tau$ as
\begin{equation}\label{a3}
dS_{5}^{2}=\left(\frac{\psi}{\psi_0}\right)^{2}\left[F(\tau,r)d\tau^{2}-J(\tau,r)\left(dr^{2}+r^{2}(d\theta^{2}
+sin^{2}\theta d\phi^{2})\right)\right]-d\psi^{2},
\end{equation}
where the metric functions $F(\tau,r)$ and $J(\tau,r)$ are given by
\begin{equation}\label{a4}
F(\tau,r)=a^{2}(\tau)\left[1-\frac{G\zeta\psi_0}{2a(\tau)r}\right]^{2}\left[1
+\frac{G\zeta\psi_0}{2a(\tau)r}\right]^{-2},\quad J(\tau,r)
=a^{2}(\tau)\left[1+\frac{G\zeta\psi_0}{2a(\tau)r}\right]^{4},
\end{equation}
with $d\tau=a^{-1}(\tau)dt$ and $a(\tau)=-1/(H\tau)$, so that the
Hubble parameter is a constant given by
\begin{equation}
H=a^{-2}\, \frac{d a}{d\tau}.
\end{equation}
As it was shown in \cite{antigravity}, for certain values of
$\zeta$ and $\psi_0$ the metric in (\ref{a1}) has two natural
horizons. The inner horizon is the analogous of the Schwarzschild
horizon and the external one is the analogous of the Hubble
horizon. In the metric (\ref{a3}), these horizons can in principle
be expressed in terms of the new dynamical coordinates.

\subsection{The 5D dynamics of the scalar field}

Now we consider a 5D massless scalar field which is free of any
interactions. The dynamics is given by the Klein-Gordon type field
equation
\begin{equation}\label{a5}
\frac{1}{\sqrt{|g_5|}}\frac{\partial}{\partial
y^{a}}\left[\sqrt{|g_5|}g^{ab}\varphi_{,b}\right]=0,
\end{equation}
where $\sqrt{|g_5|}=(\psi/\psi_0)^{4}F^{1/2}J^{3/2}r^{2}sin\theta$
is the determinant of the covariant metric tensor $g_{ab}$. On the
metric (\ref{a3}), the equation (\ref{a5}) becomes
\begin{equation}\label{a6}
\frac{1}{\sqrt{F J^3}}\frac{\partial}{\partial
\tau}\left(\sqrt{\frac{J^3}{F}}\frac{\partial\varphi}{\partial\tau}\right)-\frac{1}{2}
\left(\frac{1}{FJ}\frac{\partial F}{\partial
r}+\frac{1}{J^2}\frac{\partial J}{\partial r}\right)\frac{\partial
\varphi}{\partial r}- \frac{1}{J}\nabla^{2}\varphi
-\left(\frac{\psi}{\psi_0}\right)^{-2}\frac{\partial}{\partial
\psi}\left[\left(\frac{\psi}{\psi_0}\right)^{4}
\frac{\partial\varphi}{\partial\psi}\right]=0,
\end{equation}
where
\begin{equation}
\nabla^2 \equiv \frac{1}{r^2} \frac{\partial}{\partial r}
\left(r^2 \frac{\partial}{\partial r}\right) + \frac{1}{r^2
\sin\theta} \frac{\partial}{\partial \theta} \left( \sin\theta
\frac{\partial}{\partial \theta}\right) + \frac{1}{r^2
\sin^2\theta} \frac{\partial^2}{\partial \phi^2},
\end{equation}
is the 3D Laplacian in spherical coordinates.

Assuming that $\varphi(\tau,r,\theta,\phi,\psi)$ can be separated
in the form
\begin{equation}\label{a7}
\varphi(\tau,r,\theta,\phi,\psi)\sim
\Phi(\tau,r)G(\theta,\phi)\Omega(\psi),
\end{equation}
the expression (\ref{a6}) leaves to
\begin{eqnarray}
&&\left(\frac{\psi}{\psi_0}\right)^{-2}\frac{d}{d\psi}\left[\left(\frac{\psi}{\psi_0}\right)^{4}
\frac{d\Omega}{d\psi}\right]+M^2\Omega = 0, \label{a9a}\\
\label{a9} && \frac{1}{\sqrt{FJ}}\frac{\partial}{\partial
\tau}\left(\sqrt{\frac{J^3}{F}}\frac{\partial\Phi}{\partial\tau}\right)-\frac{1}{2}
\left(\frac{1}{F}\frac{\partial F}{\partial
r}+\frac{1}{J}\frac{\partial J}{\partial
r}\right)\frac{\partial\Phi}{\partial r}-
\frac{1}{r^2}\frac{\partial}{\partial
r}\left(r^{2}\frac{\partial\Phi}{\partial
r}\right) \nonumber \\
&& -\left(\frac{l(l+1)}{r^2}-M^2J\right)\Phi=0 \label{a10},
\end{eqnarray}
where $M$ is a separation constant with mass units and $l$ is an integer parameter. In
deriving (\ref{a10}) we have separated out the angular part of $\varphi$.

\subsection{The 4D induced field equation}

Assuming that the 5D spacetime can be foliated by a family of hypersurfaces $\Sigma:\psi=\psi_0$,
from the metric (\ref{a3})
we obtain that the 4D induced metric on every leaf $\Sigma:\psi=\psi_0$ is given by
\begin{equation}\label{b1}
dS_{4}^{2}=F(\tau,r)d\tau^{2}-J(\tau,r)[dr^{2}+r^{2}(d\theta^{2}+\sin^{2}\theta
d\phi^{2})],
\end{equation}
where the metric functions $F(\tau,r)$ and $J(\tau,r)$ can be now written in terms of the physical
mass $m=\zeta\psi_0$ (introduced by the first time in \cite{antigravity}), in the form
\begin{equation}\label{b2}
F(\tau,r)=a^{2}(\tau)\left[1-\frac{Gm}{2a(\tau)\,r}\right]^{2}\left[1+\frac{Gm}{2a(\tau)\,r}\right]^{-2},\quad
J(\tau,r)=a^{2}(\tau)\left[1+\frac{Gm}{2a(\tau)\,r}\right]^{4},
\end{equation}
valid for $r > G m /(2a)$. The induced metric (\ref{b1}) describes
a black hole in an expanding universe, where the expansion is
driven by a kind of cosmological constant, whose value in general
depends of the value of $\psi_0$, which is related to the Hubble
constant: $\psi_0=1/H$. From the equations (\ref{a6}) and
(\ref{a9a}), the 4D induced field equation on $\Sigma$, reads
\begin{equation}\label{b3}
\frac{1}{\sqrt{FJ^3}}\frac{\partial}{\partial\tau}\left[\sqrt{\frac{J^3}{F}}\frac{\partial\,{\bar
\varphi}}{\partial\tau}\right]-\frac{1}{2}
\left(\frac{1}{FJ}\frac{\partial F}{\partial
r}+\frac{1}{J^2}\frac{\partial J}{\partial
r}\right)\frac{\partial\,{\bar\varphi}}{\partial r}
-\frac{1}{J}\nabla^{2}\,\bar\varphi + M^2\,{\bar\varphi}=0,
\end{equation}
where
${\bar\varphi}(\tau,r,\theta,\phi)=\varphi(\tau,r,\theta,\phi,\psi_0)$
is the effective scalar field induced on the generic hypersurface
$\Sigma$. Here, $M$ plays the role of the mass of the inflaton
field on the 4D brane (\ref{b1}), such that the effective 4D
equation of motion for $\bar\varphi$ reads: $\left[ ^{(4)}
\square+ M^2\right] \bar\varphi =0$. This result was previously
obtained in a different manner using the fact that the 5D linear
momentum $P^a = M u^a$ is conserved on a 5D Ricci flat spacetime,
so that $P^a_{;a}=0$\cite{rindler}. Hence, as in other 5D
inflationary models, the 4D mass of the inflaton field is induced
geometrically through the foliation $\psi=\psi_0$, because on the
5D metric (\ref{a6}) the equation of motion for
$\varphi(\tau,r,\theta\phi)$ is $^{(5)} \square
\varphi(\tau,r,\theta,\phi) =0$, so that the field $\varphi$ can
be viewed in the spacetime (\ref{a6}) as a free and massless
scalar field. Furthermore, the extra dimension is related to the
BH mass $m$: $\psi_0 = m/\zeta= c^2/H$ (we use natural units
$c=\hbar=1$).
\\

On the other hand, according to \cite{antigravity} the length
scale that separates regions on which gravity changes from
attractive to repulsive, is the gravitational-antigravitational
radius, which in the coordinates $(T,R)$ is given by
\begin{equation}\label{b4}
R_{ga}=(Gm\psi^2_0)^{1/3}.
\end{equation}
However, due to the coordinate transformation (\ref{a2}), in the new coordinates $(\tau,r)$
this radius must obey the relation
\begin{equation}\label{b5}
R_{ga}=a(\tau)\,r_{ga}\left[1+\frac{Gm}{2a(\tau)\,r_{ga}}\right]^{2},
\end{equation}
where $r_{ga}$ is denoting the gravitational-antigravitational
radius in the new coordinates. Solving this equation for $r_{ga}$
we obtain
\begin{equation}\label{b6}
r_{ga}=\frac{1}{2a(\tau)}\left[R_{ga}-Gm\pm\sqrt{R_{ga}^{2}-2GmR_{ga}}\right],
\end{equation}
where the solution with the minus sign is not a physical solution. Moreover, in order to $r_{ga}$
to be a real value quantity,
we require the condition $R_{ga}^{2}-2GmR_{ga}\geq 0$ to be hold. This condition can be rewritten in the form
\begin{equation}\label{b7}
R_{ga}\geq 2Gm.
\end{equation}
Inserting (\ref{b4}) in (\ref{b7}), we obtain the restriction
\begin{equation}\label{b8}
m^{2}\leq\frac{\psi_{0}^{2}}{8G^{2}}.
\end{equation}
If we consider the foliation $\psi_0=c^2/H$ and the fact that for
$c=\hbar=1$ the Newtonian constant is $G=M^{-2}_p$, the condition
(\ref{b8}) yields
\begin{equation}\label{bb8}
\epsilon=\frac{ m H}{M^2_p} \le \frac{1}{2 \sqrt{2}} \simeq
0.353553,
\end{equation}
and hence the fifth coordinate must be restricted to the following
condition:
\begin{equation}\label{condition}
\psi_0 \geq {2 \sqrt{2} m \over M^2_p}.
\end{equation}
This condition allow us to consider objects whose mass satisfies
the parameter $\epsilon = GmH$
to be a small parameter. The same restriction has been used in \cite{CNW}
with different motivation, without affecting the range of validity of the dynamical coordinates.\\

\section{The expansion of the induced scalar field $\bar\varphi$}

We expand the induced scalar field ${\bar\varphi}$ as
\begin{equation}
\bar\varphi(\vec r,\tau) = \int^{\infty}_{0} dk\,\sum_{lm}
\left[a_{klm} \bar\Phi_{klm}(\vec r,\tau)+a^{\dagger}_{klm}
\bar\Phi^*_{klm}(\vec r,\tau)\right],
\end{equation}
where
\begin{equation}
\bar\Phi_{klm}(\vec r,\tau)= k^2 \,j_l\left(kr\right)
\bar\Phi_{kl}(\tau) Y_{lm}(\theta,\phi),
\end{equation}
where $Y_{lm}(\theta,\phi)$ are the spherical harmonics, $j_{l}(kr)$ are the spherical Bessel
functions and the annihilation and
creation operators obey the algebra
\begin{equation}
\left[a_{klm}, a^{\dagger}_{k'l'm'}\right] = \delta(k- k')
\delta_{ll'} \delta_{m m'}, \qquad \left[a_{klm},
a_{k'l'm'}\right]=\left[a^{\dagger}_{klm},
a^{\dagger}_{k'l'm'}\right]=0.
\end{equation}
Hence, once we take into account the addition theorem for spherical harmonics, we obtain for the
mean squared fluctuations
\begin{equation}\label{fluct}
\left< 0\left| \bar\varphi^2\left(\vec r,\tau\right)
\right|0\right> = \int^{\infty}_{0} \frac{dk}{k} \sum_{l}
\frac{2l+1}{4\pi} k^5 j^2_l(kr)
\left|\bar\Phi_{kl}(\tau)\right|^2.
\end{equation}
Now, if we assume that $\bar{\varphi}(\tau,r,\theta,\phi)=\bar{\Phi}(\tau,r)\bar{G}(\theta,\phi)$,
then the equation for $\bar\Phi(r,\tau)$ on the hypersurface $\Sigma$ can be written as
\begin{eqnarray}
\frac{\partial^2 {\bar\Phi_l}}{\partial\tau^2} & - &
\frac{2}{\tau} \frac{\partial {\bar\Phi_l}}{\partial\tau} -
\frac{2}{r} \frac{\partial {\bar\Phi_l}}{\partial r} -
\frac{\partial^2 {\bar\Phi_l}}{\partial r^2} - \left[\frac{l\left(
l+1\right)}{r^2} - M^2\right] {\bar\Phi_l}  \nonumber  \\
& = & \left( 1- \frac{J}{F}\right) \frac{\partial^2
{\bar\Phi_l}}{\partial\tau^2} - \left[ \frac{2}{\tau} +
\frac{1}{\sqrt{F J}} \frac{\partial}{\partial\tau}
\left(\frac{J^3}{F}\right)^{1/2} \right] \frac{\partial
{\bar\Phi_l}}{\partial\ \tau} \nonumber \\
& - & M^2 \left(J -1\right) {\bar\Phi_l} + \frac{1}{2} \left(
\frac{1}{F} \frac{\partial F_l}{\partial r} + \frac{1}{J}
\frac{\partial J}{\partial r}\right)
\frac{\partial\bar\Phi_l}{\partial r}. \label{a11}
\end{eqnarray}
Next,using the fact that $\epsilon$ is a small parameter as
indicated by (\ref{bb8}), we propose the following expansion for
$\bar\Phi_l $ in orders of $\epsilon$:
\begin{equation}\label{a13}
\bar\Phi_l(r,\tau) = \bar\Phi^{(0)}_l + \bar\Phi^{(1)}_l +
\bar\Phi^{(2)}_l+ ....
\end{equation}
If we expand the right hand side of the equation (\ref{a11}) as powers
of $\epsilon $ \cite{CNW}, we obtain
\begin{eqnarray}
&- & 8 \left(\frac{\epsilon\tau}{2 r}\right) \left[
\frac{\partial^2{\bar\Phi}^{(0)}_l}{\partial\tau^2} -
\frac{1}{\tau} \frac{\partial {\bar\Phi}^{(0)}_l}{\partial\tau} -
\frac{M^2}{2H^2\tau^2} {\bar\Phi}^{(0)}_l\right] \nonumber \\
& - & 30 \left(\frac{\epsilon \tau}{2 r}\right)^2 \left[
\frac{\partial^2  {\bar\Phi}^{(1)}_l}{\partial \tau^2} -
\frac{1}{15} \frac{1}{\tau} \frac{\partial
{\bar\Phi}^{(1)}_l}{\partial {\bar\Phi}} - \frac{M^2}{5 H^2
\tau^2}{\bar\Phi}^{(1)}_l\right] + ... \label{a12}
\end{eqnarray}
Thus, we are now able to calculate solutions for
$\bar{\Phi}_l(r,\tau)$ at zeroth and first orders in the
expansion. Higher orders require numerical analysis.

Finally, the spectrum for the squared fluctuations (\ref{fluct})
can be written using the expansion (\ref{a13}) in the following
manner
\begin{eqnarray}
{\cal P}_k(\tau) & = & \sum_{l} \frac{(2l+1)}{4\pi} k^5 \,j^2_l(k
r) \left[ \bar\Phi^{(0)}_{kl} + \bar\Phi^{(1)}_{kl} + ...\right]
\left[ \left(\bar\Phi^{(0)}_{kl}\right)^* +
\left(\bar\Phi^{(1)}_{kl}\right)^* + ... \right] \nonumber \\
& = & \frac{k^3}{2\pi^2} \left|\bar\Phi^{(0)}_{kl}\right|^2 +
\frac{H^2}{4\pi^2} \epsilon \sum^{\infty}_{l=1}\, (2l+1)\, j^2_l(k
r)\, \Delta^{(1)}_{kl} + ...\, , \label{spect}
\end{eqnarray}
such that
\begin{equation}\label{spect1}
\Delta^{(1)}_{kl} = \left(\frac{4\pi^2}{H^2 \epsilon}\right)
\frac{k^5}{4\pi} \left|\bar\Phi^{(0)}_{kl}
\left(\bar\Phi^{(1)}_{kl}\right)^* + \bar\Phi^{(1)}_{kl}
\left(\bar\Phi^{(0)}_{kl}\right)^*\right] = \frac{2\pi}{H^2
\epsilon} k^5 {\rm Re}
\left[\bar{\Phi}^{(1)}_{kl}\,\left(\bar\Phi^{(0)}_{kl}\right)^*\right].
\end{equation}

In the following subsections we shall study the spectrum at
first-order, as expanded in (\ref{spect}). We are interested
mainly in the two limit cases; the Hubble and the Schwarzschild
horizons.

\subsection{The  zeroth order expansion for the induced scalar
field}

Using the equations (\ref{a11}), (\ref{a13}) and (\ref{a12}), we
obtain that the dynamics for a massive scalar field in a spatially
homogeneous de Sitter case is described by
\begin{equation}
\frac{\partial^2 {\bar\Phi^{(0)}}_l}{\partial\tau^2} -
\frac{2}{\tau} \frac{\partial {\bar\Phi}^{(0)}_l}{\partial\tau} -
\frac{2}{r} \frac{\partial {\bar\Phi}^{(0)}_l}{\partial r} -
\frac{\partial^2 {\bar\Phi}^{(0)}_l}{\partial r^2} -
\left[\frac{l\left( l+1\right)}{r^2} - M^2 a^2(\tau)\right]
{\bar\Phi}^{(0)}_l=0,
\end{equation}
where the last term corresponds with the induced mass of the
scalar field. If we use the Bessel transform
\begin{equation}
\bar\Phi^{(0)}_l(r,\tau) =  \int^{\infty}_{0} dk\, k^2\, j_l(kr)\,
{\bar\Phi^{(0)}}_{kl}(\tau),
\end{equation}
we obtain the zeroth order dynamics for the modes
${\bar\Phi^{(0)}}_{kl}$
\begin{equation}
\frac{\partial^2 {\bar\Phi^{(0)}}_{kl}}{\partial\tau^2} -
\frac{2}{\tau} \frac{\partial {\bar\Phi^{(0)}}_{kl}}{\partial\tau}
+ \left(k^2 + \frac{M^2}{H^2 \tau^2}\right)
{\bar\Phi^{(0)}}_{kl}=0.
\end{equation}
If we use the Bunch-Davies vacuum, we obtain the normalized modes
solution
\begin{equation}
{\bar\Phi^{(0)}}_{kl} = A_1 \, \left(-\tau\right)^{3/2} {\cal
H}^{(1)}_{\nu}\left[-k\,\tau\right] + A_2 \,
\left(-\tau\right)^{3/2} {\cal
H}^{(2)}_{\nu}\left[-k\,\tau\right],
\end{equation}
where ${\cal H}^{(1,2)}_{\nu}$ are respectively the first and
second kind Hankel functions with $\nu^2 = {9 \over 4} - {M^2\over
H^2}$. The normalization constants are
\begin{equation}\label{nor}
A_2 = -\frac{\sqrt{\pi} H}{2}\,e^{-i \nu \pi/2}, \qquad A_1 = 0.
\end{equation}
If we take into account that at zero order $l$ can take only the
value $l=0$, we obtain that the spectrum of the fluctuations
without sources is
\begin{equation}
{\cal P}^{(0)}_{kl}(\tau) = \frac{k^3}{2\pi^2}
\left|\bar\Phi^{(0)}_{kl} \right|^2 = \frac{H^2}{\pi}
\left(\frac{-k \tau}{2}\right)^{3}  \, {\cal
H}^{(2)}_{\nu}[-k\tau] \, {\cal H}^{(1)}_{\nu}[-k\tau],
\end{equation}
where $k$ given by
\begin{equation}
k = \frac{2\pi}{a(\tau) r} \left[1+\frac{G m}{2 a(\tau)
r}\right]^{-2},
\end{equation}
is the wave number on a physical frame.

\subsubsection{Zeroth order: power spectrum on cosmological scales}

On cosmological scales we obtain that the following inequality is
fulfilled
\begin{equation}
\frac{G m}{2 a(\tau) r_H} \ll 1,
\end{equation}
so that the wavenumber of fluctuations at the horizon entry will
be
\begin{equation}\label{aa1}
k_H \simeq \frac{2\pi}{a(\tau) r_H} = -\frac{2\pi H\tau}{r_H},
\end{equation}
where we have made use of the fact that $a(\tau)= -1/(H\tau)$,
with $\tau \leq 0$. Furthermore, at the end of inflation $\tau
\rightarrow 0$, so that ${\cal H}^{(2)}_{\nu}[-k\tau] \simeq
{-i\over \pi} \Gamma(\nu) \left(-k\tau/2\right)^{-\nu}$. The power
spectrum on scales close to the Hubble horizon is
\begin{equation}
\left.{\cal P}^{(0)}_{kl}(\tau)\right|_{H} = \frac{k^3}{2\pi^2}
\left|\bar\Phi^{(0)}_{kl}\right|^2_{H} \simeq \left(\frac{\pi
H\tau^2}{r_H}\right)^{3-2\nu} \frac{\Gamma^2(\nu) H^2}{\pi^3},
\end{equation}
which depends on the mass of the inflaton field $M$, because
$\nu^2 = {9\over 4} - {M^2\over H^2}$. For a nearly  scale
invariant power spectrum: $\nu \simeq 3/2$, we obtain
\begin{equation}
\left.{\cal P}^{(0)}_{kl}(\tau)\right|_{H,\nu \simeq 3/2} \simeq
\frac{H^2}{4\pi^2},
\end{equation}
which is the standard result for quantum fluctuations.

\subsubsection{Zeroth order: power spectrum on Schwarzschild scales}

Another interesting limit case is the Schwarzschild scale. On these
scales the following inequality is fulfilled
\begin{equation}\label{hole}
\frac{G m}{2 a(\tau) r_{Sch}} \simeq 1,
\end{equation}
and the wavenumber of the scalar field fluctuations, when they enter to
the Schwarzschild horizon, is given by
\begin{equation}\label{aa2}
k_{Sch} \simeq \frac{8\pi a(\tau) r_{Sch}}{\left(Gm\right)^2} =
-\frac{8\pi r_{Sch}}{H\tau \left(Gm\right)^2},
\end{equation}
that increases linearly with the scale factor $a(\tau)$, so that
$k_{Sch} \tau \gg 1$, and the power of the spectrum at the end of
inflation related to the scalar field fluctuations at zero order,
is
\begin{equation}
\left.{\cal P}^{(0)}_{kl}(0)\right|_{Sch} \simeq
\left(\frac{2}{\sqrt{2}} \frac{r_{Sch}}{G m} \right)^2 \simeq
\frac{\left( H \tau\right)^2}{\pi},
\end{equation}
where we have used the expression (\ref{hole}). Notice that the
spectrum decreases dramatically as $a^{-2}$ and tends to zero at
the end of inflation (i.e., for $\tau \rightarrow 0$). A very
important fact is that the spectrum is related to the value of the
fifth coordinate (we remember that we are using natural units):
$\left.{\cal P}^{(0)}_{kl}(0)\right|_{Sch} \simeq {\tau^2\over
\pi\psi^2_0}$, which also is related to the mass, $m$, of the BH
through the condition (\ref{condition}).

\subsection{First order expansion of the induced scalar field}

The first order expansion for $\bar\Phi$ satisfies
\begin{eqnarray}
\frac{\partial^2 {\bar\Phi^{(1)}}_l}{\partial\tau^2} & - &
\frac{2}{\tau} \frac{\partial {\bar\Phi}^{(1)}_l}{\partial\tau} -
\frac{2}{r} \frac{\partial {\bar\Phi}^{(1)}_l}{\partial r} -
\frac{\partial^2 {\bar\Phi}^{(1)}_l}{\partial r^2} -
\left[\frac{l\left( l+1\right)}{r^2} -
\frac{M^2}{H^2\tau^2}\right] {\bar\Phi}^{(1)}_l\nonumber \\
&=& -4 \left(\frac{\epsilon\tau}{r}\right) \left[
\frac{\partial^2{\bar\Phi}^{(0)}_l}{\partial\tau^2} -
\frac{1}{\tau} \frac{\partial {\bar\Phi}^{(0)}_l}{\partial\tau} -
\frac{M^2}{2H^2\tau^2} {\bar\Phi}^{(0)}_l\right].\label{a14}
\end{eqnarray}
The right hand side of (\ref{a14}) can be considered as a source
term $J^{(1)}_l(r,\tau)$, of the form
\begin{eqnarray}
J^{(1)}_l(r,\tau)& = &-4 \left(\frac{\epsilon\tau}{r}\right)
\left[ \frac{\partial^2{\bar\Phi}^{(0)}_l}{\partial\tau^2} -
\frac{1}{\tau} \frac{\partial {\bar\Phi}^{(0)}_l}{\partial\tau} -
\frac{M^2}{2H^2\tau^2} {\bar\Phi}^{(0)}_l\right] \nonumber \\
& = & -4 \left(\frac{\epsilon\tau}{r}\right) \int^{\infty}_{0}
dk\,k^2\,j_l\left(kr\right) \left[ \frac{\partial^2
{\bar\Phi^{(0)}}_{kl}}{\partial\tau^2} - \frac{1}{\tau}
\frac{\partial {\bar\Phi^{(0)}}_{kl}}{\partial\tau} -
\frac{M^2}{2H^2\tau^2} {\bar\Phi^{(0)}}_{kl}\right].
\end{eqnarray}
To solve this equation we use the Green's function
$G(r,\tau;r',\tau ')$
\begin{equation}
G_l(r,\tau;r',\tau ') = \int^{\infty}_{0}
dk\,k^2\,g_k\left(\tau,\tau '\right)\, j_l\left(kr\right)
j_l\left( k r'\right),
\end{equation}
where $g_k$ satisfies the dynamics
\begin{equation}
\frac{\partial^2 g_k}{\partial\tau^2} - \frac{2}{\tau}
\frac{\partial g_k}{\partial\tau} + \left[k^2 + \frac{M^2}{H^2
\tau^2}\right] g_k=\frac{2}{\pi}\, \delta\left(\tau -
\tau'\right).
\end{equation}
The general solution for this equation is
\begin{eqnarray}
g_k(\tau,\tau') & = &  \left(-\tau\right)^{3/2} \,\left\{
A_1\,{\cal J}_{\nu}\left(-k\tau\right) + A_2\, {\cal
Y}_{\nu}\left(-k\tau\right)- \left[ \frac{{\cal
J}_{\nu}\left(-k\tau\right)}{(-\tau ')^{1/2}} {\cal
Y}_{\nu}\left(-k\tau '\right) \right.\right. \nonumber \\
& - & \left. \left. \frac{{\cal
Y}_{\nu}\left(-k\tau\right)}{(-\tau ')^{1/2}}{{\cal
J}_{\nu}\left(-k\tau '\right)}\right]\right\},
\end{eqnarray}
where $0>\tau >\tau'$. Hence, the field
$\bar\Phi^{(1)}\left(r,\tau\right)$ can be expressed using the
retarded Green function
\begin{equation}
\bar\Phi^{(1)}_l\left(r,\tau\right) = \int^{\infty}_{0}
dk\,k^2\,j_l\left(kr\right)\,\bar\Phi^{(1)}_{kl}\left(\tau\right)
= \int^{\infty} dr'\,r'^2 \int^{0}_{\tau_i}
d\tau'\,G_l(r,\tau;r',\tau ')\,J^{(1)}_l\left(r',\tau'\right),
\end{equation}
where $\tau_i$ denotes the time when the source begins to operate
and the modes $\bar\Phi^{(1)}_{kl}$ can be represented as
\begin{equation}\label{a14'}
\bar\Phi^{(1)}_{kl}\left(\tau\right) = \epsilon \left\{
\alpha_{kl}\left(\tau\right) \,\bar\Phi^{(0)}_{k}\left(\tau\right)
+
\beta_{kl}\left(\tau\right)\,\left[\bar\Phi^{(0)}_{k}\left(\tau\right)\right]^*\right\},
\end{equation}
or explicitly
\begin{eqnarray}
\bar\Phi^{(1)}_{kl}\left(\tau\right) & = &  \frac{e^{-i\nu\pi/2}
\sqrt{\pi}\,\epsilon}{2H} \left(-\tau\right)^{3/2}
\int^{\tau}_{\tau_i} d\tau'\left\{ A_1 {\cal J}_{\nu}(-k\tau) +
A_2 {\cal Y}_{\nu}(-k\tau) - \left[ {\cal J}_{\nu} (-k\tau) {\cal
Y}_{\nu}(-k\tau') - {\cal Y}_{\nu}(-k\tau) {\cal
J}_{\nu}(-k\tau') \right] \right\} \nonumber \\
& \times & \int^{\infty}_{0} \frac{dk'}{(k')^3} \left\{ {\cal
H}^{(2)}_{\nu}(-k'\tau') \left[ H^2\left(3-4(\nu+\nu^2 + (\tau'
k')^2\right) + 2 M^2\right] -4 k' H^2 \tau' {\cal
H}^{(2)}_{\nu}\left(-k'\tau'\right) \right\} \nonumber \\
& \times & \int^{\infty}_{0} dr'\,r'\,j_l(k r')\,j_l(k' r').
 \label{a15}
\end{eqnarray}
If we take into account the normalization conditions on the
background modes ${\bar\Phi^{(0)}}_{kl}$, considered in
(\ref{nor}), the expression (\ref{a15}) can be rewritten as
\begin{eqnarray}
\bar\Phi^{(1)}_{kl}\left(\tau\right) & = &
\frac{\epsilon}{H^2}\int^{\tau}_{\tau_i} d\tau'
 \left\{ H^2\,\left[\bar{\Phi}^{(0)}_{kl}(-k\tau)\right]  + \frac{i}{2} e^{-i\nu\pi/2} \, \left[
e^{-i\nu\pi/2}\left[\bar{\Phi}^{(0)}_{kl}(-k\tau)\right]^* {\cal
H}^{(2)}_{\nu} (-k\tau')\right.\right. \nonumber \\
& - & \left.\left.
e^{i\nu\pi/2}\left[\bar{\Phi}^{(0)}_{kl}(-k\tau)\right] {\cal
H}^{(1)}_{\nu} (-k\tau') \right] \right\}
 \nonumber \\
& \times & \int^{\infty}_{0} dk'(k')^2 \left\{ {\cal
H}^{(2)}_{\nu} (-k'\tau') \left[ H^2\left(3-4(\nu+\nu^2 + (\tau'
k')^2\right) + 2 M^2\right] \right. \nonumber \\
&-& \left. 4 k' H^2 \tau' {\cal H}^{(2)}_{\nu+1} (-k'\tau')
\right\}  \times \int^{\infty}_{0} dr'\,r'\,j_l(k r')\,j_l(k' r'),
 \label{a16}
\end{eqnarray}
such that
\begin{eqnarray}
\int^{1}_{0} dr'\,r'\,j_l\left(k r'\right)j_l\left(k' r'\right) &
= & \frac{\pi}{2} \frac{
(k')^l\Gamma(l+1)}{\Gamma(l+3/2)\,\Gamma(1/2)} \times
F\left(l+1,\frac{1}{2};l+\frac{3}{2},k'^2\right), \label{i1} \\
\int^{\infty}_{1} dr'\,r'\,j_l\left(k r'\right)j_l\left(k'
r'\right) & = & \frac{\pi}{2} \frac{
(k')^{-(l+2)}\Gamma(l+1)}{\Gamma(l+3/2)\,\Gamma(1/2)} \times
F\left(l+1,\frac{1}{2};l+\frac{3}{2},\frac{1}{k'^2}\right).
\label{i2}
\end{eqnarray}
The expression (\ref{i1}) is valid for $k \in (0,1)$ and the
expression (\ref{i2}) corresponds to the case $k\in (1,\infty)$.
Furthermore, $F(a,b;c,x)$ is the hypergeometric function. From
eqs. (\ref{a14'}) and (\ref{a16}), we obtain the following
expressions for the coefficients $\alpha_{kl}$ and $\beta_{kl}$:
\begin{eqnarray}
\alpha_{kl}(\tau) & = & \epsilon \int^{\tau}_{\tau_i} d\tau'
\left\{ 1- \frac{i}{2 H^2} \, {\cal H}^{(2)}_{\nu}
(-k\tau')\right\} \int^{\infty}_{0} dk'(k')^2 \nonumber \\
& \times & \left\{ {\cal H}^{(2)}_{\nu} (-k'\tau') \left[
H^2\left(3-4(\nu+\nu^2 + (\tau' k')^2\right) + 2 M^2\right] -4 k'
H^2 \tau' {\cal
H}^{(2)}_{\nu+1}(-k'\tau')  \right\} \nonumber \\
& \times & \int^{\infty}_{0} dr'\,r'\,j_l(k r')\,j_l(k' r'), \label{co1} \\
\beta_{kl}(\tau) & = & \frac{i \epsilon \,e^{-i\nu\pi}}{2 H^2}
\,\int^{\tau}_{\tau_i} d\tau'\, {\cal H}^{(1)}_{\nu} (-k\tau')
\int^{\infty}_{0} dk'(k')^2 \nonumber \\
& \times & \left\{  {\cal H}^{(2)}_{\nu} (-k'\tau') \,\left[
H^2\left(3-4(\nu+\nu^2 + (\tau' k')^2\right) + 2 M^2\right]
-4 k' H^2 \tau' {\cal H}^{(2)}_{\nu+1}(-k'\tau')  \right\} \nonumber \\
& \times & \int^{\infty}_{0} dr'\,r'\,j_l(k r')\,j_l(k' r'),
\label{co2}
\end{eqnarray}
where $\tau_i=-1/H=-\psi_0$ is the conformal time when inflation
starts, which will be considered when $a(\tau_i)=1$. The
wavenumbers related to the horizons $k_H$ and $k_{Sch}$, are given
respectively by the equations (\ref{aa1}) and (\ref{aa2}) and
$k_{ga}$ is the gravitational-antigravitational wavenumber
\begin{equation}\label{aaa3}
k_{ga} = \frac{2\pi}{a(\tau) r_{ga}} \left[\frac{2 a(\tau)
r_{ga}}{2 a(\tau) r_{ga} + Gm}\right]^2,
\end{equation}
where $r_{ga}$ is given by (\ref{b6}). Furthermore, the power
spectrum for the $\Phi^{(1)}$-fluctuations is [see eqs.
(\ref{spect}) and (\ref{spect1})]
\begin{equation}
{\cal P}^{(1)}_{kl} = \frac{H^2}{4\pi^2} \epsilon
\,\Delta^{(1)}_{kl}(\tau),
\end{equation}
where $\Delta^{(1)}_{kl}(\tau)$ is given by
\begin{equation}\label{aa3a}
\Delta^{(1)}_{kl}(\tau) = \frac{2\pi k^5}{\epsilon H^2} {\rm Re}
\left\{ \Phi^{(1)}_{kl}(\tau)
\left(\Phi^{(0)}_{kl}(\tau)\right)^*\right\} =\frac{2\pi k^5}{H^2}
\left\{ \left|\Phi^{(0)}_{kl}\right|^2 {\rm
Re}\left[\alpha_{kl}(\tau)\right] + {\rm
Re}\left[\beta^*_{kl}\left(\Phi^{(0)}_{kl}\right)^2\right]\right\},
\end{equation}
where we have used the expression (\ref{a14'}). In the following
subsections we shall calculate the spectrums
$\Delta^{(1)}_{kl}(\tau)$ on scales where gravity is repulsive
(big scales, of today cosmological scales) and attractive (small
scales, or today astrophysical scales) in the first order
approximation.

\subsubsection{First order: power spectrum on cosmological scales}

Now we consider the spectrum (\ref{aa3a}). To calculate
$\alpha_{kl}(\tau)$ and $\beta_{kl}(\tau)$, we shall make use of
the asymptotic expressions for the first and second kind Hankel
functions: ${\cal H}^{(1,2)}_{\nu}[x] \simeq \mp {i\over \pi}
\Gamma(\nu) \left(x/2\right)^{-2}$ [in our case $x(\tau) = -k\tau
\ll 1$], in (\ref{co1}) and (\ref{co2})
\begin{eqnarray}
\left.\alpha_{kl}(\tau)\right|_{-k\tau\ll 1} & \simeq &
\frac{i\,\epsilon \pi}{2} \frac{\Gamma(l+1)}{\Gamma(l+3/2)
\Gamma(1/2)} \int^{k_{ga}}_{k_H} dk' \, \int^{\tau}_{\tau_i}
d\tau'\,\left\{1+\frac{2^{\nu-1} k^{-\nu} \Gamma(\nu)
\left(-\tau'\right)^{-2\nu}}{\pi H^2}\right\}
\left(k'\right)^{2-\nu+l}   \nonumber \\
& \times & \left\{ 2^{\nu} \frac{\Gamma(\nu)}{\pi} \left[ H^2
\left[3-4(\nu+\nu^2 + (k'\tau')^2 + 2 M^2\right]\right] \right.
\nonumber \\
&-& \left. \frac{k' \tau'}{\pi} H^2 \Gamma(\nu+1)
\left(k'\right)^{-(\nu+1)} 2^{\nu+3}
\left(-\tau'\right)^{-(\nu+1)}\right\} \times  _2F_1[l+1,1/2;l+3/2;\left(k'\right)^2], \nonumber \\\label{ee1} \\
\left.\beta_{kl}(\tau)\right|_{-k\tau\ll 1} &\simeq & i
\frac{\epsilon \Gamma(l+1) \Gamma^2(\nu) e^{-i\nu\pi}}{4\pi H^2
\Gamma(l+3/2) \Gamma(1/2)} \int^{\tau}_{\tau_i} d\tau'
\left(-\frac{k \tau'}{2}\right)^{-\nu} \nonumber
\\
& \times & \int^{k_{ga}}_{k_H} dk' \, \left(k'\right)^l \,\,
_2F_1[l+1,1/2;l+3/2;
\left(k'\right)^2] \nonumber \\
& \times & \left\{ \left[ H^2 \left[3-4(\nu+\nu^2 + (\tau' k')^2
\right]+ 2 M^2\right] - 4 i\frac{k'\tau' H^2\Gamma(\nu+1)}{\pi}
\left(-\frac{k'\tau´}{2}\right)^{-(\nu+1)}\right\}, \nonumber
\\ \label{ee2}
\end{eqnarray}
where $_2F_1[a,b;c;y(k)]$ is the hypergemetric function. When the
horizon entry we obtain the following approximated spectrum on
cosmological scales:
\begin{equation}\label{aa4}
\left.\Delta^{(1)}_{kl}(\tau_e)\right|_{-k\tau\ll 1} \simeq 2 {\rm
Re}\left[ \left.\alpha_{kl}(\tau_e)\right|_{-k\tau\ll 1} -
\left.\beta_{kl}(\tau_e)\right|_{-k\tau\ll 1}\right],
\end{equation}
where $\left.\alpha_{kl}(\tau_e)\right|_{-k\tau\ll 1}$ and
$\left.\beta_{kl}(\tau_e)\right|_{-k\tau\ll 1}$ are the
expressions (\ref{ee1}) and (\ref{ee2}) when $\tau \rightarrow
\tau_e$, with $1/(-\tau_e) > H e^{60}$. In other words $(-\tau_e)$
is sufficiently close to zero to preserve the required large scale
flatness of the universe at the end of inflation. In this limit
case the coefficients are
\begin{eqnarray}
\left.\alpha_{kl}(\tau_e)\right|_{-k\tau\ll 1} & \simeq &
\frac{i\,\epsilon \pi}{2} \frac{\Gamma(l+1)}{\Gamma(l+3/2)
\Gamma(1/2)} \int^{k_{ga}}_{k_H} dk' \, \int^{\tau_e}_{\tau_i}
d\tau'\,\left\{1+\frac{2^{\nu-1} k^{-\nu} \Gamma(\nu)
\left(-\tau'\right)^{-2\nu}}{\pi H^2}\right\}
\left(k'\right)^{2-\nu+l} \nonumber \\
& \times & \left\{2^{\nu} \frac{\Gamma(\nu)}{\pi} \left[ H^2
\left[3-4(\nu+\nu^2 + (k'\tau')^2 + 2 M^2\right]\right] - \frac{
k' \tau'}{\pi} H^2 \Gamma(\nu+1) \left(k'\right)^{-(\nu+1)}
2^{\nu+3} \left(-\tau'\right)^{-(\nu+1)}\right\}\nonumber \\
&\times &  _2F_1[l+1,1/2;l+3/2;\left(k'\right)^2], \label{eea1} \\
\left.\beta_{kl}(\tau_e)\right|_{-k\tau\ll 1} &\simeq & i
\frac{\epsilon \Gamma(l+1) \Gamma^2(\nu) e^{-i\nu\pi}}{4\pi H^2
\Gamma(l+3/2) \Gamma(1/2)} \int^{\tau_e}_{\tau_i} d\tau'
\left(-\frac{k \tau'}{2}\right)^{-\nu} \int^{k_{ga}}_{k_H} dk' \,
\left(k'\right)^l \,\, _2F_1[l+1,1/2;l+3/2;
\left(k'\right)^2] \nonumber \\
& \times & \left\{ \left[ H^2 \left[3-4(\nu+\nu^2 + (\tau' k')^2
\right]+ 2 M^2\right] - 4 i\frac{k'\tau' H^2\Gamma(\nu+1)}{\pi}
\left(-\frac{k'\tau'}{2}\right)^{-(\nu+1)}\right\}. \label{eea2}
\end{eqnarray}
Since
\begin{equation}
2 {\rm Re}\left[ \left.\alpha_{kl}(\tau_e)\right|_{-k\tau\ll
1}\right]=0,
\end{equation}
the function
\begin{eqnarray}
\left.\Delta^{(1)}_{kl}(\tau_e)\right|_{-k\tau\ll 1} & \simeq &
\frac{\epsilon \,2^{2(\nu+1)} \Gamma(l+1) \Gamma^2(\nu)
\Gamma(\nu+1)}{\pi^2 \Gamma(l+3/2) \Gamma(1/2)} \, k^{-\nu}
\nonumber \\
& \times & \int^{\tau_e}_{\tau_i} d\tau'\,(-\tau')^{-2\nu} \,
\int^{k_{ga}}_{k_H} dk' (k')^{l-\nu}\,_2F_1[l+1,1/2;l+3/2;
\left(k'\right)^2],
\end{eqnarray}
is determined by real part of (\ref{eea2}) which describes the
spectrum of the scalar field fluctuations at the end of inflation
in the infrared region, in which gravity is repulsive. This
spectrum is enclosed by the wavenumbers $k_{ga}$ and $k_H$,
related respectively to the gravitational-antigravitational radius
and the Hubble horizon. Notice that this spectrum is linear with
$\epsilon=mH/M^2_p$, so that its amplitude depends on the BH mass
$m$. In absence of the BH (i.e., for $m=0$),
$\Delta^{(1)}_{kl}=0$.

\subsubsection{First order: power spectrum on small scales}

To calculate the spectrum at first order on small scales, we shall
make use of the asymptotic expressions for big argument Hankel
functions: ${\cal H}^{(1,2)}_{\nu}[x(\tau)] \simeq  \left(2/(\pi
x)\right)^{1/2} e^{\pm i [x(\tau)-\nu\pi/2-\pi/4]}$, where
$x(\tau) = -k\tau$. The coefficients $\alpha_{kl}(\tau)$ and
$\beta_{kl}(\tau)$ are given by
\begin{eqnarray}
\left.\alpha_{kl}(\tau)\right|_{-k\tau\gg 1} & \simeq & \epsilon
\sqrt{\frac{\pi}{2}} \frac{\Gamma(l+1)}{\Gamma(l+3/2) \Gamma(1/2)}
\int^{\tau}_{\tau_i} \left\{ 1 - \frac{i}{H^2}\sqrt{\frac{2}{\pi}}
\left(-k \tau'\right)^{-1/2} \,e^{-
i [-k\tau'-\nu\pi/2-\pi/4]}\right\} \nonumber \\
& \times & \int^{k_{Sch}}_{k_{ga}} \left(k'\right)^{-l}\,
_2F_1\left[l+1,1/2;l+3/2;1/(k')^2\right]\, \left\{ \left(-k'
\tau'\right)^{-1/2} e^{- i
[-k'\tau'-\nu\pi/2-\pi/4]}\right.  \nonumber \\
& \times & \left.\left[ H^2 \left[3-4(\nu+\nu^2 + (\tau'
k')^2\right] + 2 M^2\right] - 4 k' \tau' H^2 \,\left(-k'
\tau'\right)^{-1/2}\, e^{- i
[-k'\tau'-(\nu+1)\pi/2-\pi/4]}\right\}, \nonumber \\
&& \\
\left.\beta_{kl}(\tau)\right|_{-k\tau\gg 1} & \simeq & \frac{i
\epsilon}{2 H^2} \frac{\Gamma(l+1)}{\Gamma(l+3/2) \Gamma(1/2)}
e^{-i\nu\pi} \int^{\tau}_{\tau_i} d\tau'\, \left(-k
\tau'\right)^{-1/2}\, \,e^{- i [-k\tau'-\nu\pi/2-\pi/4]}
\int^{k_{Sch}}_{k_{ga}} dk' (k')^{-l}\, \nonumber
\\
& \times &  \left\{ \left(-k' \tau'\right)^{-1/2}\,e^{- i
[-k'\tau'-\nu\pi/2-\pi/4]} \left[ H^2 \left[ 3-4(\nu+\nu^2 +
(\tau' k')^2\right] +2 M^2\right]\right.
\nonumber \\
& - & \left. 4 \sqrt{\frac{2}{\pi}} H^2 k'\tau' \left(-k'
\tau'\right)^{-1/2} \,e^{- i
[-k'\tau'-(\nu+1)\pi/2-\pi/4]}\right\}\,
_2F_1\left[l+1,1/2;l+3/2;1/(k')^2\right]. \nonumber \\
\end{eqnarray}

The function $\left.\Delta^{(1)}_{kl}(\tau_e)\right|_{-k\tau\gg
1}$ describes the spectrum of the scalar field fluctuations on
scales between the the Schwarzschild horizon and
gravitational-antigravitational distances, which are related to a
wavenumber $k_{ga}$, and describes the region of the spectrum
where gravity is attractive. At the end of inflation
$\left.\Delta^{(1)}_{kl}(\tau_e)\right|_{-k\tau\gg 1}$ is given by
\begin{equation}\label{e0}
\left.\Delta^{(1)}_{kl}(\tau_e)\right|_{-k\tau\gg 1} = \frac{2\pi
k^5}{H^2} \left\{ \left|\Phi^{(0)}_{kl}\right|^2_{-k\tau \gg 1}
{\rm Re}\left[\left.\alpha_{kl}(\tau_e)\right|_{-k\tau\gg
1}\right] + {\rm
Re}\left[\left.\beta_{kl}(\tau_e)\right|_{-k\tau\gg
1}\left(\Phi^{(0)}_{kl}\right)^2_{-k\tau \gg 1}\right]\right\},
\end{equation}
where
\begin{eqnarray}
\left.\alpha_{kl}(\tau_e)\right|_{-k\tau\gg 1} & \simeq & \epsilon
\sqrt{\frac{\pi}{2}} \frac{\Gamma(l+1)}{\Gamma(l+3/2) \Gamma(1/2)}
\int^{\tau_e}_{\tau_i} \left\{ 1 -
\frac{i}{H^2}\sqrt{\frac{2}{\pi}} \left(-k \tau'\right)^{-1/2}
\,e^{-
i [-k\tau'-\nu\pi/2-\pi/4]}\right\} \nonumber \\
& \times & \int^{k_{Sch}}_{k_{ga}} \left(k'\right)^{-l}\,
_2F_1\left[l+1,1/2;l+3/2;1/(k')^2\right]\, \left\{ \left(-k'
\tau'\right)^{-1/2} e^{- i
[-k'\tau'-\nu\pi/2-\pi/4]}\right.  \nonumber \\
& \times & \left.\left[ H^2 \left[3-4(\nu+\nu^2 + (\tau'
k')^2\right] + 2 M^2\right] - 4 k' \tau' H^2 \,\left(-k'
\tau'\right)^{-1/2}\, e^{- i
[-k'\tau'-(\nu+1)\pi/2-\pi/4]}\right\}, \nonumber \\
& & \\
\left.\beta_{kl}(\tau_e)\right|_{-k\tau\gg 1} & \simeq &
\frac{i \epsilon}{2 H^2} \frac{\Gamma(l+1)}{\Gamma(l+3/2)
\Gamma(1/2)} e^{-i\nu\pi} \int^{\tau_e}_{\tau_i} d\tau'\, \left(-k
\tau'\right)^{-1/2}\, \,e^{- i [-k\tau'-\nu\pi/2-\pi/4]}
\int^{k_{Sch}}_{k_{ga}} dk' (k')^{-l}\, \nonumber
\\
& \times &  \left\{ \left(-k' \tau'\right)^{-1/2}\,e^{- i
[-k'\tau'-\nu\pi/2-\pi/4]} \left[ H^2 \left[ 3-4(\nu+\nu^2 +
(\tau' k')^2\right] +2 M^2\right]\right.
\nonumber \\
& - & \left. 4 \sqrt{\frac{2}{\pi}} H^2 k'\tau' \left(-k'
\tau'\right)^{-1/2} \,e^{- i
[-k'\tau'-(\nu+1)\pi/2-\pi/4]}\right\}\,
_2F_1\left[l+1,1/2;l+3/2;1/(k')^2\right], \nonumber \\
&& \\
\left|\Phi^{(0)}_{kl}\right|^2_{-k\tau \gg 1} & \simeq &  \frac{H^2}{2\pi } \frac{ (-\tau)^2}{k}, \\
\left(\Phi^{(0)}_{kl}\right)^2_{-k\tau \gg 1} & \simeq &
\frac{H^2}{2\pi } \frac{ (-\tau)^2}{k}\, e^{-2i\left[(-k\tau)
-\pi/4\right]},
\end{eqnarray}
which also becomes zero in absence of the BH. The calculation of
$\left.\Delta^{(1)}_{kl}(\tau_e)\right|_{-k\tau\gg 1}$ is very
complicated, but one can see clearly using the expansion for the
hypergeometric function:
\begin{displaymath}
_2F_1(a,b;c;z)=\frac{\Gamma(c)}{\Gamma(a) \Gamma(b)}
\sum_{n=0}^{\infty} \frac{\Gamma(a+n)\, \Gamma(b+n)}{\Gamma(c+n)}
\frac{z^n}{n!},
\end{displaymath}
that terms of $k'$ with positive potentia are relevant for large
values of $l$, i.e., for smaller scales.

\section{Final comments}

In this letter we have studied inflationary quantum scalar field
fluctuations of an effective 4D scalar field, in the framework of
an extended version of general relativity derived from a 5D vacuum
theory of gravity. The interesting aspect of this extended version
of general relativity, is that gravity manifests itself as
repulsive at large (cosmological) scales and attractive at short
scales \cite{antigravity}. We assume a 5D spacetime  described
geometrically by the SdS metric (\ref{a1}), endowed with a 5D
massless scalar field $\varphi$. In order to consider an
inflationary cosmological setting we write the metric in
(\ref{a1}) in the dynamical coordinate chart
$(\tau,r,\theta,\phi)$. In this new coordinates the length scale
that separates regions on which gravity changes from attractive to
repulsive, named the gravity-antigravity radius $r_{ga}$, becomes
dynamical and it varies inversely proportional to the conformal
scale factor $a(\tau)$, as it was shown in the expression
(\ref{b6}). This can be interpreted on 4D cosmological settings as
when the universe expands, the region where gravity is repulsive
becomes larger every time. When we go down from five to four
dimensions via a foliation of the 5D spacetime in the fifth
coordinate, it is a well known result that the 5D massless scalar
field $\varphi$ can be seen in 4D as a massive effective scalar
field, as it is shown in this particular case for
$\bar{\varphi}(\tau,r)$ by the equation (\ref{b3}). This is an
important characteristic of 5D free scalar fields were they move
on an effective 4D hypersurface obtained by a static foliaton [in
our case $\psi = \psi_0 = c^2/H=m/\zeta$].

We quantized the 4D effective scalar field $\bar{\varphi}$
following the canonical procedure and we found that its quantum
modes evolve according to the equation (\ref{a11}). Given the
difficulty of finding exact solutions for the modes equation
(\ref{a11}), we decided to use an expansion for the modes
$\bar{\Phi}(\tau,r)$ in powers of a small parameter
$\epsilon=mH/M_{p}^{2}$. The fact that $\epsilon \leq 0.353553$ is
a small parameter arises naturally from the model and provides a
constraint for the extra coordinate: $\psi_0 \geq (2 \sqrt{2} m)/ M^2_p$.
From our analysis we obtained that it is possible to
have exact solutions $\bar{\Phi}$ at least at zeroth and first
orders in the expansion. Finally, one important result obtained in
this model is that at the end of an stage of de-Sitter inflation,
the corresponding spectrum of fluctuations
\begin{itemize}
\item at zeroth order results to be independent of the scalar
field mass $M$ on Schwarzschild scales, \item while on
cosmological scales it exhibits a mass $M$ dependence through the
parameter $\nu^2 = {9 \over 4} - {M^2\over H^2}$, because the
spectrum go as $k^{3-2\nu}$.
\end{itemize}
At first order of expansion, we found that on both length scales
the spectrum depends on the inflaton field mass $M$ and the
amplitude depends on the BH mass $m$, which is the source of
spatial inhomogeneity in our model.

\section*{Acknowledgements}

\noindent J.E.M.A acknowledges CONACYT (M\'exico) and M.B.
acknowledges UNMdP and CONICET (Argentina) for financial support.


\bigskip

\begin{thebibliography}{99}
\bibitem{einasto} J. Einasto, {\em et. al}, Astroph. J. {\bf 519}:
441 (1999).
\bibitem{Wbook} J. M. Overduin and P. S. Wesson, Phys. Rept. {\bf
283}: 303 (1997); P.S. Wesson, Space-Time-Matter, World Scientific, Singapore (1999).
\bibitem{Maartens} Roy Maartens, Livng Rev. Rel. {\bf 7}, 7 (2004).
\bibitem{Nor} G. Nordstr$\ddot{o}$m, Phys. Z. {\bf 15}, 504 (1914).
\bibitem{KKT} T. Kaluza, Sitz. Preuss. Akad. Wiss. {\bf 33}, 996 (1921); O. Klein, Phys. Z {\bf 37} 895, (1926).
\bibitem{BT} M. Pavsic, The Landscape of Theoretical Physics: A Global View- From Point Particles
to the Brane World and Beyond in the Search of a Unifying Principle of Physics. Springer Heidelberg (2002).
\bibitem{stm} P. S. Wesson, Gen. Rel. Grav. {\bf 16}: 193 (1984);
P. Wesson, Gen. Rel. Grav. {\bf 22}: 707 (1990); P. S. Wesson,
Phys. Lett. {\bf B276}: 299 (1992); P. S. Wesson and J. Ponce de
Leon, J. Math. Phys. {\bf 33}: 3883 (1992); H. Liu and P. S.
Wesson, J. Math. Phys. {\bf 33}: 3888 (1992); P. Wesson, H. Liu
and P. Lim, Phys. Lett. {\bf B298}: 69 (1993).
\bibitem{Equi} J. Ponce de Leon, Mod. Phys. Lett. {\bf A16}, 2291-2304, (2001).
\bibitem{2} P. S. Wesson, J. Ponce de Leon,
J. Math. Phys. {\bf 33}: 3883 (1992).
\bibitem{CMT1} C. Romero, R. Tavakol and R. Zalaletdinov, Gen. Rel. Grav. {\bf 28}: 365 (1996).
\bibitem{11} H. Y. Liu and B. Mashhoon, Phys. Lett. {\bf A272}:
26 (2000); \\
B. Mashhoon and P. S. Wesson, Class. Quant. Grav. {\bf 21}: 3611
(2004).
\bibitem{12} M. L. Liu, H. Y. Liu, L. X. Xu and P. S.
Wesson, Mod. Phys. Lett. {\bf A21}: 39 (2006).
\bibitem{rs} L. Randall and R. Sundrum, Phys. Rev. Lett. {\bf 83}:
3370 (1999);\\
L. Randall and R. Sundrum, Phys. Rev. Lett. {\bf 83}: 4690 (1999).
\bibitem{antigravity} J. E. Madriz Aguilar and M. Bellini, Phys. Lett. {\bf B679}: 306 (2009).
\bibitem{planar} T. Shiromizu, D. Ida and T, Torii, J. High Energy Physics {\bf 11}: 010 (2001).
\bibitem{rindler} W. Rindler, {\em Essential Relativity}. (2nd.
edition) Springer, Berlin, (1977); \\
S. S. Seahra and P. S. Wesson, Gen. Rel. Grav. {\bf 33}, 1731
(2001);\\
P. S. Wesson, {\em Five-dimensional Physics, Classical and Quantum
Consequences of Kaluza-Klein Cosmology}, World Scientific, New
Jersey (2006).
\bibitem{npb} M. Bellini, Nucl. Phys. {\bf B660}: 389 (2003).
\bibitem{CNW} H. T. Cho, K. W. Ng and I, C. Wang, {\em Scalar
field fluctuations in Schwarzschild-de Sitter space-time}. E-print
arXiv: 0905.2041.

\end{thebibliography}
\end{document}